\documentclass[aps,superscriptaddress,amsmath,amssymb,twocolumn,amsfonts]{revtex4-1}
\usepackage{amsmath}
\usepackage{graphicx}
\graphicspath{{Pictures/}}
\usepackage{hyperref}
\hypersetup{pdfnewwindow=true, colorlinks=true, linkcolor=magenta, anchorcolor=blue, citecolor=red, filecolor=blue, menucolor=blue, urlcolor=blue}
\usepackage{cleveref}
\usepackage{color}
\usepackage{float}
\usepackage{soul}
\usepackage{ulem}
\usepackage{braket}

\newcommand{\vk}{\ensuremath{\mathbf{k}}}
\newcommand{\dr}{\ensuremath{\mathbf{r}}}
\newcommand{\vq}{\ensuremath{\mathbf{q}}}

\begin{document}

\title{Intrinsic Hallmarks of Phonon-Induced Charge Order in Cuprates}

\author{S. Banerjee}
\affiliation{Theoretical Physics III, Center for Electronic Correlations and Magnetism, Institute of Physics, University of Augsburg, 86135 Augsburg, Germany,}
\author{W. A. Atkinson}
\affiliation{Department of Physics and Astronomy, Trent University, Peterborough, Ontario K9L 0G2, Canada}
\author{A. P. Kampf}
\affiliation{Theoretical Physics III, Center for Electronic Correlations and Magnetism, Institute of Physics, University of Augsburg, 86135 Augsburg, Germany,}

\date{\today}
\begin{abstract}
Charge-density wave (CDW) modulations in underdoped high-temperature cuprate superconductors remain a central puzzle in condensed matter physics. However, despite a substantial experimental verification of this ubiquitous phase in a large class of high $T_{\mathrm{c}}$ cuprates, a complete theoretical explanation of this phase is still missing. Here, we build upon our recent proposal that the CDW in underdoped cuprates (Y- and Bi- based compounds) emerges from a unique cooperation of the B$_{1g}$ bond-buckling phonon with strong electronic correlations. We assume a static mean-field lattice distortion with B$_{1g}$ symmetry, regardless of its origin, with a commensurate wave vector $\vq^*=(2\pi/3,0)/(0,2\pi/3)$. We show that such a phonon-induced CDW (both uni- and biaxial) reconstructs the Fermi surface, leading to electron and hole pockets, with relevant quantum oscillation frequencies in close consistency with the experiments. Furthermore, a systematic analysis of the symmetry of the intra-unit-cell charge modulations on the copper-oxygen planes is provided. We find that the atomic charge modulation on the CuO$_2$ unit cell is predominantly of $s$-wave character \--- in support of the recent experimental observation.
\end{abstract}

\maketitle

\section{Introduction}

The charge-density wave\cite{Comin} in the copper-oxygen planes of hole-doped cuprates\cite{Ghiringhelli821,Wu2011,Chang2012,Tabis:2014,PhysRevLett.110.137004,Dasilva,Peng2018} has been affirmed as one ubiquitous component of their phase diagram. Although the existence of CDW correlations is well established, both the underlying mechanism and the detailed structure of the charge-ordered state are poorly understood. State-of-the-art numerical calculations in two-dimensional $t$-$J$\cite{PhysRevLett.113.046402} and Hubbard models~\cite{Huang1161,Zheng1155} do yield tendencies for charge- or spin-stripe formation. Yet, the opposite trends of the CDW wave vector variation with doping in different cuprate materials raise doubts about a universal mechanism. Furthermore, the associated orbital symmetry of the charge modulation remains undetermined. It is commonly argued that the charge is redistributed between the oxygen atoms in the planar CuO$_2$ unit cell and hence described by a $d$-wave form factor~\cite{PhysRevB.82.075128,PhysRevLett.111.027202,Efetov2013,PhysRevB.76.020511,PhysRevB.77.094504,Atkinson_2015,FujitaE3026,Comin2015}. This is supported by scanning tunneling microscopy (STM) experiments, which point to a predominant $d$-wave form factor modulation for the CDW state in Bi-2212~\cite{FujitaE3026}. On the contrary, resonant x-ray scattering studies of the 214 cuprates such as La$_{1.875}$Ba$_{0.125}$CuO$_4$ (LBCO) indicate a predominant $s'$-wave form factor.\cite{Achkar2016}  Whether a similar $s$-wave or the $d$-wave form factor dominates in the charge-ordered state in underdoped YBa$_2$Cu$_3$O$_{6+x}$ (YBCO) remains a subtle issue. Indeed, contrary to the earlier results recent x-ray measurements~\cite{mcmahon2019orbital} in YBCO claimed evidence for a dominant $s$-wave form factor arising from a charge redistribution on the copper sites. This new result calls for a re-examination of theoretical proposals for the CDW structure in YBCO.

An immediate consequence of the charge order is the reconstruction of the electronic Fermi surface (FS). Such reconstruction generically leads to the appearance of both electron and hole pockets, which manifest themselves as distinct frequencies in quantum oscillation (QO) experiments~\cite{Allais2014}. Biaxial charge order has previously been shown to yield nodal electron pockets,\cite{Ramshaw2011,Sebastian_2012} consistent with QO frequencies on  underdoped YBCO~\cite{Sebastian,luoista,LeBoeuf2007}, although even a uniaxial CDW can give rise to electron-like pockets if combined with a nematic distortion of the underlying lattice~\cite{PhysRevB.84.012507,PhysRevB.76.220503}. Evidence for the hole pockets associated with biaxial order is more tenuous, with the reported QOs being an order of magnitude weaker than for the electron pockets~\cite{Doiron-Leyraud2015}.  As one possible reason for the difficulty of observing these hole pockets, it has been suggested that they might be more susceptible to disorder or order parameter fluctuations~\cite{Allais2014}. In fact, direct evidence for FS reconstruction is elusive and has not been observed in Angle Resolved Photoemission Spectroscopy (ARPES)~\cite{Kivelson14395}.

Here, we build upon our recent proposal~\cite{banerjee2019phononinduced} that the CDW in underdoped YBCO arises from a unique cooperation of the B$_{1g}$ bond-buckling phonon mode~\cite{Forgan2015} with strong electronic correlations. A key element of this proposal is that the electron-phonon (\textit{el-ph}) matrix element $\mathsf{g}(\vq; \vk)$ is strongly momentum-dependent and thereby selects a unique axial ordering wave vector. Below, we show that the momentum dependence of $\mathsf{g}(\vq;\vk)$ also matters for both the orbital symmetry of the CDW and the FS reconstruction. We discuss the qualitative features of the reconstructed FS for both  uni- and biaxial charge order. Consistent with previous results,~\cite{PhysRevB.97.195153} we find hole pockets for the uniaxial CDW and both electron and hole pockets for the biaxial CDW. We notice that the relative weight of the copper and oxygen orbitals on the unreconstructed FS primarily dictates the overall distribution of elementary symmetry components of the incipient charge modulation, whereas the B$_{1g}$ phonon coupling plays a secondary role. To demonstrate this, we further provide a comparative analysis with respect to the conventional Holstein phonon coupling. 

In this paper we perform a mean-field analysis for which the order parameter is the magnitude of the static mean-field lattice distortion ($\xi_\vq$) from a softened $B_{1g}$ lattice vibration. This choice is motivated by x-ray diffraction experiments in YBCO~\cite{Forgan2015}  which found that the static CDW lattice distortion has a predominantly B$_{1g}$ character. The distortion is not purely B$_{1g}$, indicating that other modes contribute to the CDW; however, for simplicity we restrict our analysis to the dominant B$_{1g}$ component. In order to obtain quantitative estimates, we adopt the experimental value\cite{Forgan2015} for the out-of-plane oxygen displacements $\xi_{\vq} \sim 2.65 \times10^{-3}$ {\AA}, which leads to a characteristic energy scale of $\sim 14$ meV for the assumed phonon-induced CDW state. This value has been extracted from the largest displacement quoted in Table I from the resonant x-ray scattering experiment by Forgan et al.~\cite{Forgan2015}. We obtain the corresponding amplitudes for the intra-unit cell charge modulations on oxygen and copper atoms, and find values that are comparable to those inferred from NMR measurements~\cite{Kharkov2016,Wu2011}. Estimates for the QO frequencies and their sensitivity to the specific model parameters are discussed in the subsequent sections. Previously~\cite{banerjee2019phononinduced} we argued that our phonon-based mechanism is applicable to Y- or Bi- based materials, but not to 214-compounds \textit{viz.} LBCO, owing to the opposite trend of charge ordering wave vector vs. doping, along with the apparent reduction of four-fold rotation symmetry due to a subtle low-temperature tetragonal (LTT) structural phase in 214-cuprates. Following our previous proposal, we stress that the quantitative estimates provided here are relevant for YBCO or Bi-2212.

This paper is organized as follows: In Sec.~\ref{sec.Model}, we introduce the YBCO specific \textit{el-ph} Hamiltonian~\cite{banerjee2019phononinduced}. We analyze the model on the mean-field level for uniaxial and biaxial CDW states in Sec.~\ref{sec:unidir} and Sec.~\ref{sec:bidir}, respectively, and provide the estimates for the associated QO frequencies for YBCO. The classification of the orbital symmetry for the emergent charge modulation is analyzed in detail in Sec.~\ref{sec:symmetry} followed by the conclusion in Sec.~\ref{sec:conclu}.

\section{Model}\label{sec.Model}

We start from an effective three-band model for a single CuO$_2$ plane in YBCO in terms of the copper $3d_{x^2-y^2}$ and oxygen $p_x$ and $p_y$ orbitals. The downfolding procedure that generates such an effective model is provided in Appendix.~\ref{sec:downfolding}. We focus on the anti-bonding band, which couples to the out-of-plane B$_{1g}$ vibrations of the oxygen atoms through the local electric field $E_z$~\cite{PhysRevB.51.505,PhysRevLett.93.117004}. The corresponding Hamiltonian for the antibonding band is 
\begin{equation}\label{eq.el-ph}
\scalebox{0.75}[1]{$\mathcal{H}  = \sum_\vk \varepsilon_\vk c^{\dagger}_\vk c_\vk + \sum_{\vq,\vk} \mathsf{g}(\vq;\vk) c^{\dagger}_{\vk+\vq}c_\vk   \left( a_\vq + a^{\dagger}_{-\vq} \right) +\hbar\Omega_{\mathrm{P}}\sum_{\vq}a^{\dagger}_\vq a_\vq$},
\end{equation}
where $c^{\dagger}_\vk$ is the creation operator for the anti-bonding electrons with dispersion $\varepsilon_\vk$, $a_\vq$ annihilates a dispersionless B$_{1g}$ phonon mode with frequency $\hbar\Omega_{\mathrm{P}} \sim 40$ meV. The irrelevant spin degrees of freedom are dropped. Here, we assume a linear \textit{el-ph} coupling, which is justified by the small value  of the oxygen displacements, $\xi_{\bf q}$~\cite{Forgan2015}.  Previous theoretical work~\cite{Li_2015} has shown that non-linear \textit{el-ph} interactions limit CDW correlations when lattice displacements are large; however, this does not appear to be relevant here.

We adopt all the tight-binding parameters entering the dispersion $\varepsilon_\vk$ from Andersen et al.~\cite{ANDERSEN19951573}. The detailed structure of the momentum-dependent \textit{el-ph} coupling is given \cite{banerjee2019phononinduced} as $\mathsf{g}(\vq;\vk) = \gamma \tilde{\mathsf{g}}(\vq;\vk)$, where $\gamma$ is the coupling strength and 
\begin{equation}\label{ph-k}
\tilde{\mathsf{g}}(\vq;\vk) = e^x_\vq \phi_x(\vk')\phi_x(\vk) + e^y_\vq\phi_y(\vk')\phi_y(\vk),
\end{equation}
with $\vk' = \vk +\vq$. The eigenfunctions $\phi_{x,y}$ signify the orbital content of the oxygen $p_{x,y}$ orbitals in the anti-bonding band. The \textit{el-ph} coupling strength is given as $\gamma = eE_z\sqrt{\hbar/2m\Omega_{\mathrm{P}}}$, where $m$ is the mass of the oxygen atoms. We adopt the electric field value $eE_z = 3.56$ eV/{\AA} from Ref.~\onlinecite{PhysRevB.82.064513} which leads to $\gamma \sim 0.22$ eV~\cite{PhysRevB.51.505,PhysRevLett.93.117004}. The eigenvectors for the B$_{1g}$ mode are $e^{x,y}_\vq = \mp\cos(q_{y,x}/2)/M_\vq$, where the normalization factor $M_\vq = \sqrt{\cos^2 (q_x/2) + \cos^2 (q_y/2)}$.

Previously, we showed that the correlated \textit{el-ph} model gives rise to local charge fluctuations consistent with experimental observations~\cite{banerjee2019phononinduced}.  The B$_{1g}$ phonon was found to be too weak to generate true long-range order by itself (and indeed only softens by a few percent~\cite{banerjee2019phononinduced}), however charge density fluctuations will necessarily be pinned by crystalline disorder, generating static short-range charge density correlations.  Here, we simply assume that there is a static B$_{1g}$ distortion $\xi_\vq$, regardless of its origin, and the Hamiltonian in Eq.~\ref{eq.el-ph} is thus reformulated as 
\begin{equation}\label{eq.mfhamil}
\scalebox{0.73}[1]{$\mathcal{H}_{\text{MF}} = \sum_{\vk} \varepsilon_{\vk} c^{\dagger}_{\vk} c_{\vk} + \sum_{\vk, \vq^*} eE_z \mathsf{\tilde{g}}(\vq^*;\vk) \xi_{\vq^*}c^{\dagger}_{\vk +\vq^*}c_{\vk} + \frac{m\Omega_{\mathrm{P}}^2}{2}\sum_{\vq^*} |\xi_{\vq^*}|^2$},
\end{equation}
where $\xi_{\vq} = \sqrt{\hbar/2m\Omega_{\mathrm{P}}} \Braket{a_{\vq} + a^{\dagger}_{-\vq}}$, with the four possible axial wave vectors $\pm \vq^*$, $\pm \overline{\vq}^*$ oriented either along the $x$- and/or the equivalent $y$- direction. As discussed in the introduction, the various x-ray scattering experiments in YBCO suggest a doping-dependent incommensurate ordering wavevector at $(q_{\text{co}},0)/(0,q_{\text{co}})$ with $q_{\text{co}} \sim 0.3$--$0.34$ reciprocal lattice units~\cite{PhysRevLett.110.137004,Ghiringhelli821,Chang2012}. For simplicity, we assume a commensurate CDW with $\vq^* =  (\frac 13,0)2\pi $ and $\overline{\vq}^* = (0,\frac 13)2\pi$, which corresponds to a charge modulation with a periodicity of three lattice constants.  This value is close to the range $q_\text{co} = 0.323$--$0.328$ reciprocal lattice units obtained for YBCO$_{6.54}$.\cite{Forgan2015} Furthermore, the distinction between commensurate and incommensurate wavevectors is unimportant when the CDW potential is weak: incommensurate values of $q_\text{co}$ produce a cascade of shadow bands of progressively higher order in the CDW potential, while a period-3 CDW includes shadow bands up to second order only.  As we show below, this distinction is unobservable for realistic values of the CDW potential.  

\section{Uniaxial CDW \label{sec:unidir}}

\begin{figure*}[t]
\centering
\includegraphics[width=1\linewidth]{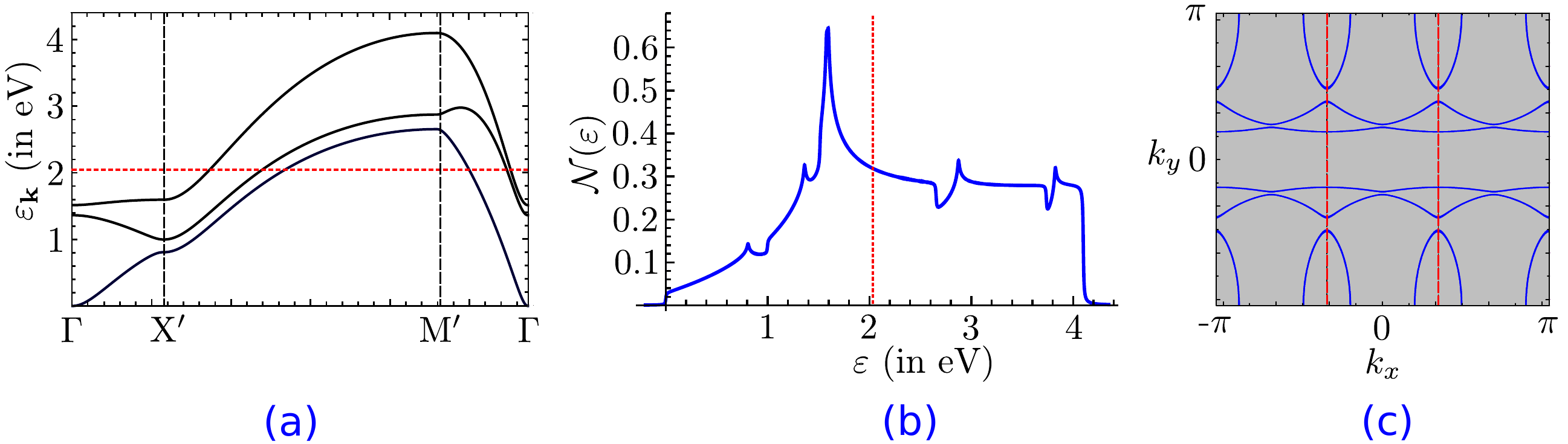} 
\caption{(a) The uniaxial CDW band structure obtained by diagonalizing the Hamiltonian in Eq.~\ref{eq.unimfhamil2} for $10 \%$ hole doping along the high-symmetry directions in the reduced Brillouin zone (RBZ) (for $\Delta_0 \sim 0.4$ eV). The corresponding chemical potential is highlighted by the dashed horizontal line. The high-symmetry points X$'$, M$'$ are defined as X$'= (\pi/3,0)$ and M$'=(\pi/3,\pi)$. (b) The modified density of states in the CDW phase. (c) The reconstructed FS in the CDW phase at the same hole doping. The dashed vertical lines specify the RBZ boundaries for the ordering wave vector $\vq^* = (2\pi/3,0)$. The RBZ contains one hole pocket at the boundaries $(\pm \pi/3,\pi)$ and $(\pm\pi/3,- \pi)$.}\label{fig:Fig1}
\end{figure*}

In this section, we analyze the reconstruction of the electronic FS for uniaxial charge order. For this case, the mean-field Hamiltonian in Eq.~\ref{eq.mfhamil} can be expressed in terms of a three-component spinor $\Psi_\vk = (c_\vk,\; c_{\vk+\vq^*},\; c_{\vk-\vq^*})^{\text{T}}$ as 
\begin{equation}\label{eq.unimfhamil1}
\mathcal{H}_{\text{MF}} = \sum_{\vk \in \text{RBZ}} \Psi^{\dagger}_\vk \underset{\sim}{\text{H}}(\vk) \Psi_\vk,
\end{equation}
where the reduced Brillouin zone (RBZ) is defined by $|k_x| \le \pi/3$ and the Hamiltonian matrix $\underset{\sim}{\text{H}}(\vk)$ is 
\begin{equation}\label{eq.unimfhamil2}
\underset{\sim}{\text{H}}(\vk) = \begin{pmatrix}
\varepsilon_{\vk} & \Delta(\vq^*;\vk) & \Delta(-\vq^*;\vk) \\
\Delta(\vq^*;\vk) & \varepsilon_{\vk+\vq^*} & \Delta(\vq^*;\vk+\vq^*) \\
\Delta(-\vq^*;\vk) & \Delta(\vq^*;\vk+\vq^*) & \varepsilon_{\vk-\vq^*}
\end{pmatrix},
\end{equation}
where $\Delta(\vq^*;\vk) = \Delta_0\tilde{\mathsf{g}}(\vq^*;\vk)$ and $\Delta_0 = eE_z \xi_{\vq^*}$ denotes the overall strength of the CDW order parameter. For future reference, we denote the eigenvalues and eigenvectors of $\underset{\sim}{\text{H}}(\vk)$ by $\varepsilon_\vk^\alpha$ and $[\psi_\alpha(\vk),\, \psi_\alpha(\vk+\vq^\ast),\, \psi_\alpha(\vk-\vq^\ast)]^T$, respectively, with $\alpha = 1,2,3$ labeling the CDW bands.

\subsection{Qualitative features \label{sec:quality}}

Throughout this section, the dispersion $\varepsilon_{\bf k}$ is taken from downfolded band-structure calculations for YBCO.\cite{ANDERSEN19951573}  Previously, we showed that the renormalization of this dispersion by strong correlations is an important key to understanding the formation of the CDW.\cite{banerjee2019phononinduced} However, once the CDW has formed, the effects of this renormalization are quantitative and can be neglected for qualitative discussions. Furthermore, for illustrative purposes we take at this stage an inflated value for the B$_{1g}$ eigenmode displacement, $\xi_\vq = 0.1$~{\AA} (implying $\Delta_0 = 0.4$ eV). The consequent band structure and the FS are shown in Figs.~\ref{fig:Fig1} (a) and (c) for a hole doping $p=0.10$, where $p=1-2n$ is measured relative to the half-filled band, and 
\begin{equation}
n = \frac 1N \sum_{\vk \in \mathrm{RBZ}} \sum_{\alpha = 1}^3f(\varepsilon^{\alpha}_\vk),
\end{equation}
with $f(x)$ the Fermi-Dirac distribution function. The chemical potential in the CDW phase is readjusted to preserve the hole doping $p$.
\begin{figure}[b]
	\centering
	\includegraphics[width=1.0\linewidth]{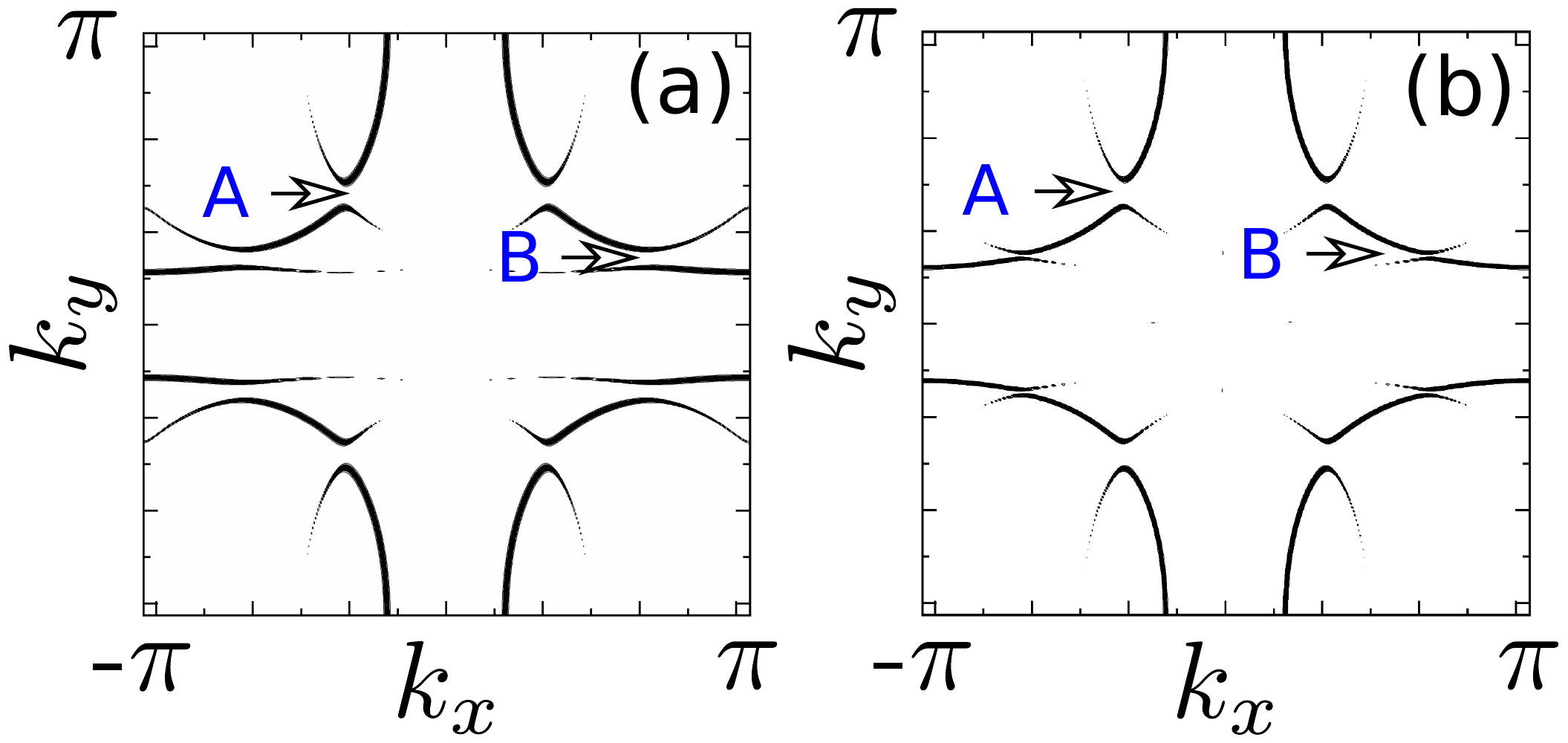} 
	\caption{Spectral function at the Fermi energy in the CDW phase for (a) the Holstein and (b) the B$_{1g}$ phonon models.  Points labeled ``A'' and ``B'' are hotspots where the Fermi surface is reconstructed by the CDW.  The CDW energies are chosen to be $\Delta_0 \approx 0.12$ eV and $\Delta_0 = 0.40$ eV in (a) and (b), respectively, which gives identical values for the band splitting at point A.  Results are for $p=0.10$.} 
	\label{fig:Fig_Holstein}
\end{figure}

Because the CDW is chosen to be commensurate, with a period of three unit cells, the CDW spectrum in Fig.~\ref{fig:Fig1}a contains three separate bands, which gives rise to the FS in Fig.~\ref{fig:Fig1}c. Notably, there is a single hole pocket centered around $(\pi/3,\pi)$, which arises from the lowest CDW band, while the open FS sheets arise from the other two bands.~\cite{PhysRevB.97.195153,PhysRevB.90.174503}. The corresponding density of states (DOS) $\mathcal{N}(\varepsilon)$, shown in Fig.~\ref{fig:Fig1}b, contains a primary Van Hove singularity characterized by the vanishing of the group velocity $\nabla_\vk \varepsilon_\vk$ (presence of saddle points in the band structure) which arises already from the unreconstructed band structure~\cite{PhysRevB.95.054518}. The additional peaks and dips in the DOS appear because of the reconstruction; each peak is accompanied by a significant suppression in the DOS. In Ref.~\onlinecite{PhysRevB.97.125147}, it was shown that such features arise at energies where bands with antiparallel group velocities are connected by $\vq^*$.

The electronic structure illustrated in Fig.~\ref{fig:Fig1} qualitatively resembles previous results for uniaxial CDWs.\cite{Sebastian_2012}  Here, we show that the matrix elements $\mathsf{g}({\bf q}; {\bf k})$ have a subtle but important effect on the FS reconstruction. Figure~\ref{fig:Fig_Holstein} compares the spectral function at the Fermi energy,
\begin{equation}\label{eq.spec}
A(\vk,\varepsilon_{\mathrm{F}}) = -\text{Tr}\;[ \text{Im}\, \underset{\sim}{G}(\vk,\varepsilon_{\mathrm{F}}) ],
\end{equation}
where $\underset{\sim}{G}(\vk,\varepsilon_{\mathrm{F}}) = \left(\varepsilon_{\mathrm{F}}- \underset{\sim}{\text{H}}(\vk) + i \delta \right)^{-1}$, for two models:  the B$_{1g}$ phonon model, and the Holstein phonon model for which $\mathsf{g}({\bf q}; {\bf k})$ is momentum independent.  For this purpose, the CDW amplitude in the Holstein model is adjusted such that the band splitting at one of the FS hotspots (labeled "A" in Fig.~\ref{fig:Fig_Holstein}) is the same in both models. Please note that the FS hotspot is defined by the portions of FS which are reconstructed by the CDW. 
	
Figure~\ref{fig:Fig_Holstein} illustrates two important points.  First, while the Fermi surfaces for both models are topologically the same as in Fig.~\ref{fig:Fig1}c, the distribution of spectral weight depends on $\mathsf{g}(\vq;\vk)$, and the Fermi surface backfolding is much less apparent in the B$_{1g}$ phonon model than in the Holstein model. Second, although the splitting at hotspot A is adjusted to be the same in both models, the band splitting at hotspot B in Fig.~\ref{fig:Fig_Holstein} is a factor of $\sim 5$ smaller for the B$_{1g}$ phonon than for the Holstein model.  The reconstruction and the spectral-weight redistribution at the secondary hotspot is thus by far less evident for the B$_{1g}$ model than for the Holstein model.  We will show below that this has important implications for the observability of the hole pocket in quantum oscillation experiments.
 
\subsection{Quantitative estimates \label{sec:estimate}}

For quantitative comparisons to experiments, we adjust our model parameters specifically to underdoped YBCO. Adopting the measured value of the B$_{1g}$ displacement $\xi_\vq \sim 2.65 \times 10^{-3}$ {\AA, we obtain $\Delta_0 = eE_z \xi_{\vq}\sim 14$ meV. Furthermore, we use a phenomenological fit ($\varepsilon^{\text{R}}_\vk$) to the measured ARPES dispersion for optimally doped Bi-2212~\cite{PhysRevB.75.184514}. Here, we rely on the one-band fitted dispersion available for the optimally doped Bi-2212, as opposed to YBCO due to the experimental difficulty to obtain a clean surface for the latter. $\varepsilon^{\text{R}}_\vk$ yields an electronic FS which closely resembles the FS from the downfolded DFT band structure used in the previous section, however the Fermi velocity is almost three times smaller~\cite{banerjee2019phononinduced}. 

The consequent reconstructed CDW FS, and the associated spectral function are shown in Fig.~\ref{fig:Fig2}a,b, respectively, at $p=0.10$ hole doping. The FS reconstruction is not apparent at all in the figure because of the small size of $\Delta_0$, and the shadow bands generated by backfolding are unobservably faint. Higher-order shadow bands, which appear when $\vq^*$ is incommensurate, are orders of magnitude weaker than the main band. The band splittings at the hotspots are 21~meV for hotspot A and 3.6~meV for hotspot B. The magntitude of the splitting at both A and B is greater than the energy resolution of ARPES experiments~\cite{RevModPhys.75.473}, and could in principle be resolved experimentally. However, they might well be masked in experiments on Bi-based cuprates by a large residual broadening at low temperature\cite{Yamasaki_2007} that is presumably due to disorder.

\begin{figure}[t]
\centering
\includegraphics[width=1.0\linewidth]{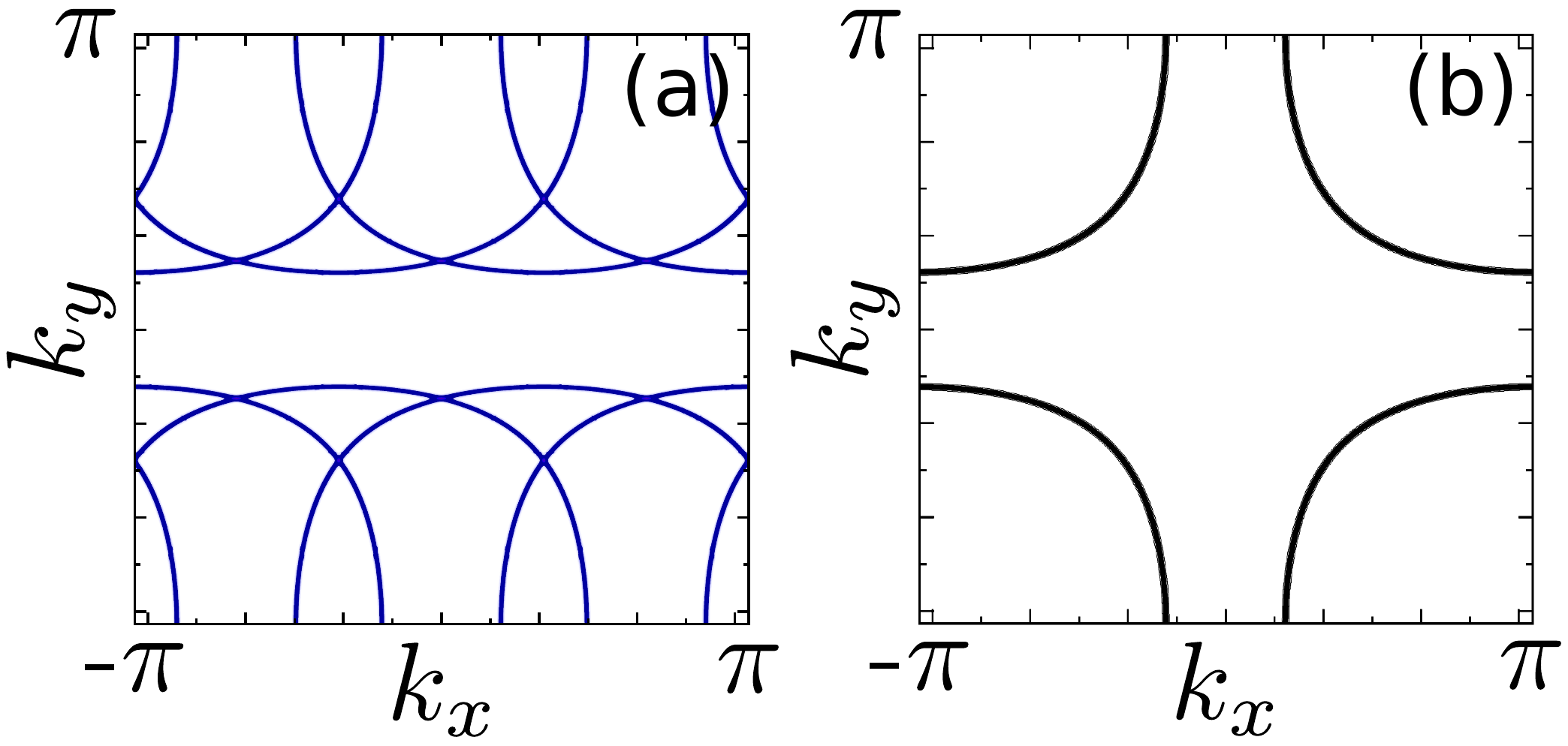} 
\caption{(a) The CDW Fermi surface for $10 \%$ hole doping obtained by diagonalzing $\underset{\sim}{\text{H}}(\vk)$ with the fitted dispersion $\varepsilon^{\text{R}}_\vk$~\cite{PhysRevB.75.184514} and the measured static $B_{1g}$ displacement $\xi_{\vq^*}$~\cite{Forgan2015}. (b) The spectral function in the CDW phase for the same parameters.} \label{fig:Fig2}
\end{figure}
 Previously we have argued~\cite{banerjee2019phononinduced} that a preference for either uni- or biaxial character of the CDW in Y- or Bi-based cuprates cannot be decided upon within the correlated \textit{el-ph} mechanism. Although it is established that QO experiments~\cite{Sebastian,Sebastian_2012} predominantly measure electron pockets (which naturally occur in the case of biaxial CDW reconstruction, see Sec.~\ref{sec:bidir}), we nevertheless analyze the relevant frequency, also for the uniaxial case. The hole pocket for the uniaxial CDW is centered around the corners of the RBZ and has an area $A \sim 1.84/a_0^2$ ($a_0$ is the lattice constant), which translates into a frequency $F = (\hbar c/2\pi e) A \sim 1260$ T. This number is to be contrasted with the observed $F_{\text{expt}} = 530 T$ in biaxial charge ordered YBCO~\cite{Doiron-Leyraud2007,PhysRevLett.104.086403,Riggs2011}. However, since, the back folded hole pockets are completely washed out as in Fig.~\ref{fig:Fig2}b, they should remain invisible to ARPES~\cite{RevModPhys.78.17}.

\subsection{Symmetry of the charge modulation: YBa$_{2}$Cu$_{3}$O$_{6+x}$ \label{sec:symmetry}}

In this section, we examine the microscopic electronic density pattern on the Cu- and the O-atoms in the copper-oxygen planes of  YBCO. The commensurate CDW wave vector $\vq^* = (2\pi/3,0)$ naturally leads to an overall modulation of the charges in the elementary CuO$_2$ unit cell with a periodicity of three lattice constants. This feature is shown in Fig.~\ref{fig:Fig5} for a single CDW unit cell, which contains three elementary CuO$_2$ unit cells. To analyze the intra-unit-cell (IUC) modulation of the charges on the Cu- and the O-atoms we first describe the elementary symmetry decomposition. There are three possibilities \--- (i) a density modulation on the Cu atoms with no modulation on the oxygen sites ($s$-wave), (ii) a uniform density modulation on the oxygen atoms only ($s’$-wave), and (iii) the opposite modulation on the O$_x$ and O$_y$ atoms with inactive Cu atoms ($d$-wave). 

Based on these three elementary patterns, we write the periodic charge modulations on the orbitals with wave-vector $\vq^*$ as~\cite{FujitaE3026,Hamidian2016,Kharkov2016}
\begin{align}\label{symmetry_decomp}
\delta n_d(\dr) & = A_s \cos\big[ \vq^*\cdot \dr + \varphi_s \big], \\
\delta n_{x}(\dr) & = A_{s'} \cos\big[ \vq^* \cdot \dr + \varphi_{s'} \big] + A_{d} \cos\big[ \vq^* \cdot \dr + \varphi_{d} \big],\nonumber \\
\delta n_{y}(\dr) & = A_{s'} \cos\big[ \vq^* \cdot \dr + \varphi_{s'} \big] - A_{d} \cos\big[ \vq^* \cdot \dr + \varphi_{d} \big], \nonumber
\end{align}
where $A_\mu$ and $\varphi_\mu$ ($\mu = s, s', d$) denote the amplitudes and phases of the individual patterns, and $\delta n_i(\dr)$ ($i=x,y,d$) denotes the charge modulation of the $i$-th orbital in unit cell $\dr$.

The CDW Hamiltonian, Eq.~(\ref{eq.unimfhamil2}) describes only the anti-bonding band, and to obtain orbitally resolved electron densities, we require the Bloch eigenfunctions $\phi_i(\vk)$ of the original three-orbital model (Appendix~\ref{sec:downfolding}). Then, the projection of the CDW onto the $i$th orbital is
\begin{equation} \label{eq:dni}
\delta n_i(\dr) = \frac 1N \sum_{\vk,\vk'} \phi_i(\vk')^\ast \phi_i(\vk) \langle c^\dagger_{\vk'} c_{\vk} \rangle e^{i(\vk-\vk')\cdot \dr}
\end{equation}
where $c_\vk$ and $c^\dagger_\vk$ are the anti-bonding band fermion operators defined before, $\phi_i(\vk)$ is the amplitude of the orbital contribution to the anti-bonding band, and $\vk' = \vk \pm \vq^\ast$~~\cite{banerjee2019phononinduced}. The expectation value is obtained from the eigenvalues $\varepsilon^{\alpha}_\vk$ and eigenvectors $\psi_\alpha(\vk)$ of the CDW Hamiltonian, Eq.~(\ref{eq.unimfhamil2})
\begin{equation}\label{eq.orb.density}
\langle c^\dagger_{\vk'} c_{\vk} \rangle = \sum_{\alpha=1}^3 \psi_\alpha(\vk') \psi_\alpha(\vk)^{\ast} f(\varepsilon^{\alpha}_\vk).
\end{equation}
The amplitudes of the various orbital symmetry components resulting from Eq.~\ref{eq.orb.density} are summarized in Table~\ref{table:symmetry}. For both the Holstein and the B$_{1g}$ phonon model, these results are calculated using the unrenormalized band structure and the inflated value $\Delta_0 = 0.4$ eV, and serve to illustrate the effects of $\mathsf{g}(\vq^\ast; \vk)$ on the CDW form factor. For both the cases, we find slight mismatch between the individual phases $\varphi_s,\varphi_{s'}, \varphi_d$ while performing the numerical fitting. However, for the purpose of this paper, we neglect this and consider a simplifying situation where all the phases are locked \textit{i.e.} $\varphi_s =\varphi_{s'}=\varphi_d = 0$. The final column shows results for the B$_{1g}$ phonon model using the fitted dispersion $\varepsilon^{\text{R}}_\vk$ from Sec.~\ref{sec:estimate}.  For comparison, the symmetry components of the lattice distortion are also included.  There are several important qualitative points to be made about these results.  
\begin{table}[t]
\setlength{\tabcolsep}{0.5em}
	\begin{tabular}{|c|c|c|c||c|}
	\hline
		& Lattice & Holstein & B$_{1g}$ & Quantitatve \\
		& (\AA) &  &  & Estimate  \\
		\hline
		$A_s$ & 0 & $4.73\times 10^{-2}$ & $1.75\times 10^{-2}$ & $1.97\times 10^{-3}$ \\
		$A_{s'}$ & 0 & $0.71 \times 10^{-2}$ & $0.01\times 10^{-2}$ & $0.07\times 10^{-3}$\\
		$A_{d}$ & 0.1 & $0.85 \times 10^{-2}$ & $0.62\times 10^{-2}$ & $0.65\times 10^{-3}$\\
		\hline
	\end{tabular}
\caption{Symmetry components of the density wave.  The first column indicates the symmetry, while the third and fourth columns list the amplitudes of the density wave components within either the B$_{1g}$ phonon or the Holstein model. The second column shows the amplitudes of the lattice distortion which has purely $d$-wave character. The amplitudes of the density-wave components in the third and fourth column are calculated using the unrenormalized band structure as in Sec.~\ref{sec:quality}. The last column contains the results obtained for the B$_{1g}$ phonon model with the renormalized band dispersions and the experimental value of $\Delta_0 \sim 14$ meV~\cite{Forgan2015}, as used in Sec.~\ref{sec:estimate}.  
The results are in units of electrons per orbital.}
\label{table:symmetry}
\end{table}

First, the pattern of the electron density does not simply follow the $B_{1g}$ symmetry of the lattice distortion. This is possible because the conventional symmetry components $s$, $s'$, and $d$ are not irreducible representations of the lattice at finite $\vq$.  Within the Holstein model, the difference between the lattice-distortion and the charge pattern comes from the orbital wavefunctions in Eq.~(\ref{eq:dni}) alone, which determine the projection of the charge modulation onto the different orbitals. A sizable $d$-orbital character on the FS leads to a significant charge modulation on the Cu sites, measured by $A_s$.  

Second, the comparison of the Holstein and the B$_{1g}$ phonon model shows that the \textit{el-ph} matrix element $\mathsf{g}(\vq^\ast;\vk)$ also affects the relative sizes of the different symmetry components.  Thus, $A_{s'}$ is significantly reduced in the B$_{1g}$ phonon model, while $A_s$ and $A_d$ are comparable in both models. The B$_{1g}$ lattice distortion has a pure $d$-wave character at the BZ center, and may therefore be expected to induce only a $d$-wave form factor to the incipient charge modulation without $s$- or $s'$ amplitudes. However, the finite CDW wavevector $\vq^*$ admixes all the symmetry component.  The reason that still we obtain a considerably smaller amplitude for $A_{s'}$ is possibly connected to the overall oxygen-orbital content variation on the FS. Whereas, the dominant amplitude $A_s$ in all the cases in Table~\ref{table:symmetry} is a result of a sizable copper orbital content.

Third, although the bandwidth renormalization enhances the CDW amplitude~\cite{banerjee2019phononinduced}, this is more than offset by the small value of $\Delta_0 = 14$~meV.  As is apparent from the last column of Table~\ref{table:symmetry}, the overall charge modulation is weak, of order 
\begin{equation}\label{eq.doping.conc.}
\delta n = A_s+2A_{s'} \sim 2.8 \times 10^{-3}
\end{equation}
electrons per unit cell.  In hindsight, the weak charge modulation could have been anticipated from the weakness of the Fermi surface reconstruction that accompanies the CDW.

\begin{figure}[t]
\centering
\includegraphics[width=1\linewidth]{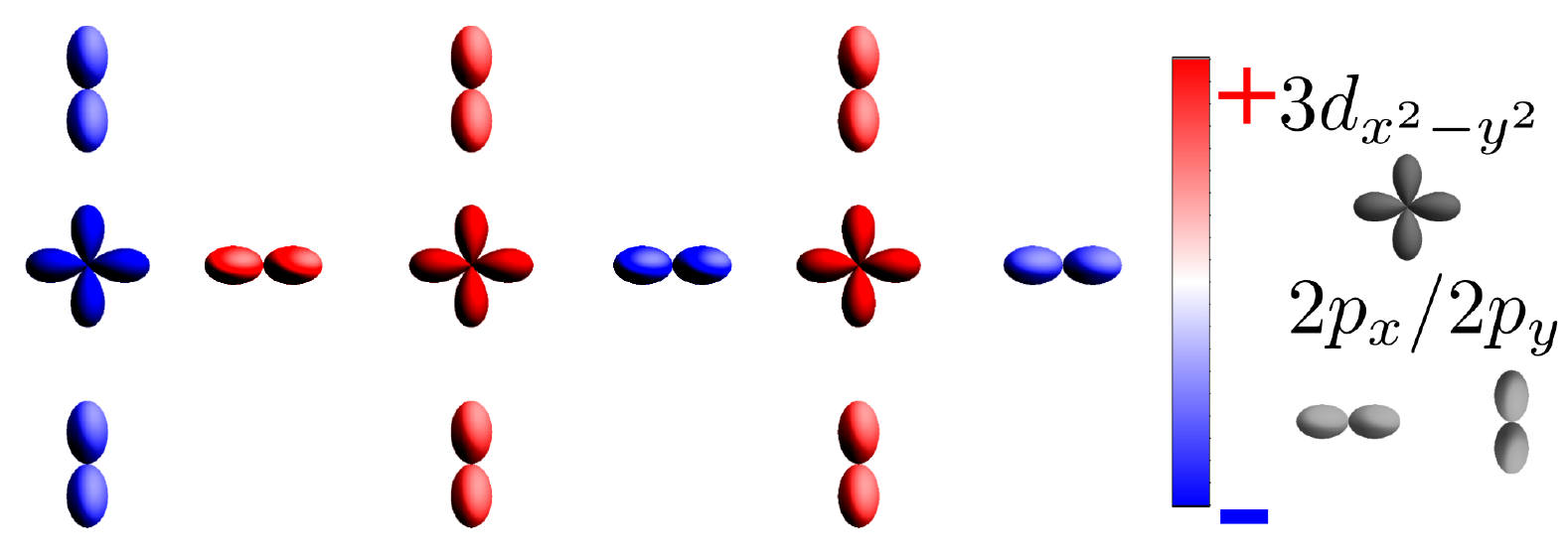} 
\caption{The modulation of the charge in the copper and oxygen orbitals due to uniaxial charge order mediated by the coupling of the electrons to the bond-buckling phonon modes of the oxygen atoms. Positive and negative variations are shown in red and blue, respectively.} \label{fig:Fig5}
\end{figure}
\begin{figure*}[t]
\centering
\includegraphics[width=1\linewidth]{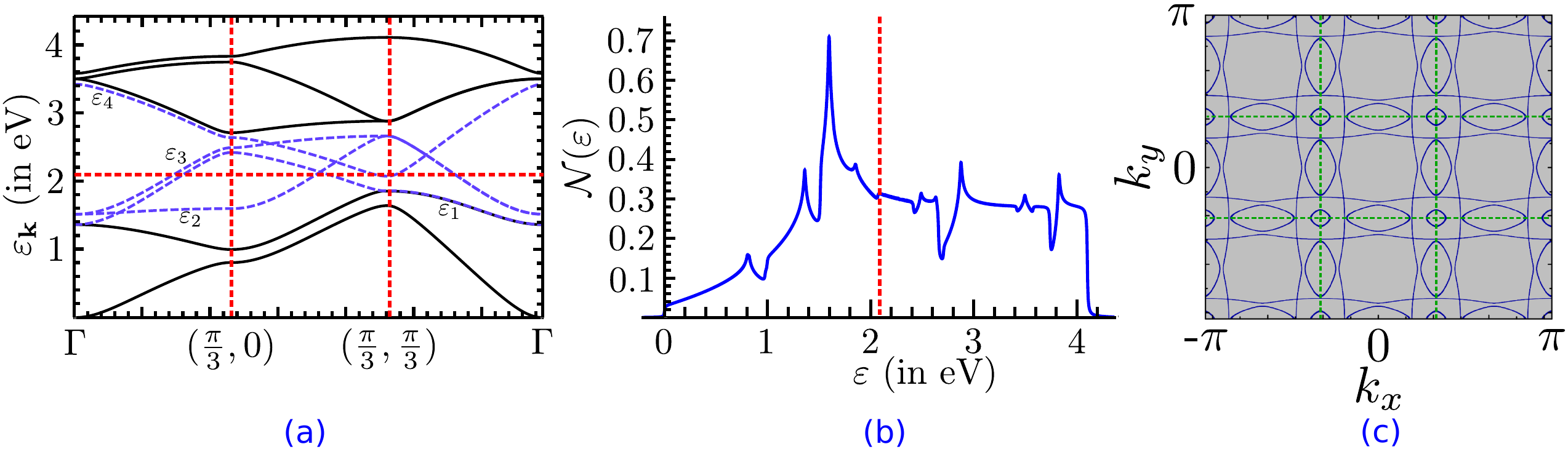} 
\caption{(a) The biaxial CDW band structure  along the high-symmetry directions in the RBZ obtained by diagonalizing the Hamiltonian in Eq.~\ref{appeq.3}. Results are for $10 \%$ hole doping and $\Delta_0 \sim 0.4$ eV. The corresponding chemical potential is highlighted by the dashed horizontal line. (b) The modified density of states in the CDW phase. (c) The reconstructed FS in the CDW phase at the same hole doping. The green dashed lines specify the RBZ boundaries for the two simultaneous ordering wave-vectors $\vq^*= (2\pi/3,0)$ and $\overline{\vq}^* = (0,2\pi/3)$. The RBZ contains two hole pockets at $(\pm\pi/3,0), (0,\pm \pi/3)$ and one electron pocket at $(\pm \pi/3, \pm\pi/3)$.} \label{fig:Fig3}
\end{figure*}

The first two of the observations above highlight the sensitivity of the charge pattern to the orbital wavefunctions $\phi_i(\vk)$.  The \textit{el-ph} matrix element $\mathsf{g}(\vq^\ast; \vk)$ is proportional to the square of the oxygen wavefunctions (Eq.~\ref{ph-k}), as is the projection of the charge modulation onto the oxygen orbitals (Eq.~\ref{eq:dni}).  Broadly speaking, a 20\% increase in the oxygen-orbital content of the FS states, i.e. in $\phi_{x,y}(\vk)$, which might be anticipated from strong correlations in the Cu $d$-orbital, would increase the overall amplitude of the charge modulation by 44\%, and double the charge modulations on the oxygen sites [$(1.2)^4 \cong 2.1$].  Given the strong sensitivity of the oxygen charge modulations to the orbital wavefunctions, the values of $A_{s'}$ and $A_{d}$ must be considered as coarse lower bounds and the true values may very well be a factor of 2 or 3 larger. Conversely, the estimate for the Cu charge modulation is much less sensitive to the orbital wavefunctions, because any enhancement of $\mathsf{g}(\vq^\ast;\vk)$ by a shift of spectral weight onto the oxygen sites will be offset by a decrease in the projection of the CDW onto the Cu site.

Despite this uncertainty, our results may be compared with an analysis of NQR linewidths in YBCO by Kharkov and Sushkov.\cite{Kharkov2016}   The authors obtained the constraints $|A_s + 0.23 A_{s'}| = 2.0\times 10^{-3}$ from the copper NQR lines, and $2.1 \times 10^{-3} < [A_{s'}^2 + A_{d}^2]^{1/2} < 6.1 \times 10^{-3}$ from the oxygen NQR lines.  Our result for $A_{s}$ is consistent with the first constraint, but our values for $A_{s'}$ and $A_{d}$ are at least a factor of 3 smaller than suggested by the second constraint. However, one must be cautious when comparing to Kharkov and Sushkov results, because the analysis in Ref.~\onlinecite{Kharkov2016} is performed with the assumption of a stripe-like (1D) CDW phase, whereas x-ray experiments are suggestive of a biaxial CDW~\cite{Forgan2015}. We expect that Kharkov and Sushkov~\cite{Kharkov2016} would consistently obtain lower amplitudes of the charge modulations, if they considered a 2D CDW.

Finally, we note that the most striking prediction of our calculation, namely that the s-wave component of the CDW is largest, is in accordance with the recent x-ray measurement on YBCO\cite{mcmahon2019orbital} in which the data for the CDW state were analyzed in favor of a dominant $s$-wave component with a sub-dominant $d$-wave form factor.

\section{Biaxial CDW\label{sec:bidir}}

In a similar fashion, we now outline the reconstruction of the electronic FS in the biaxial CDW state. Within the mean-field picture of the softened  $B_{1g}$ phonons, the quasiparticles of the CDW state are described by Eq.~\ref{eq.mfhamil} subject to the periodic modulation of the lattice distortions with both wave vectors $\vq^*=(2\pi/3,0)$ and $\overline{\vq}^* = (0,2\pi/3)$. As a result, the reduced Brillouin zone in this case is determined by $|k_x|, |k_y| \le \pi/3$. The CDW Hamiltonian in this RBZ becomes a $9\times 9$ matrix in terms of a nine-component spinor $\Psi_\vk$ (see Appendix~\ref{sec:bidirection}). For a qualitative analysis, we again take the inflated value of the B$_{1g}$ displacement $\xi_\vq \sim 0.1$ {\AA} (implying $\Delta_0 \sim 0.4$ eV), as in the analysis of the uniaxial CDW in Sec.~\ref{sec:quality}.

\subsection{Qualitative features \label{sec.quality_biaxial}}

\begin{figure}[b]
\centering
\includegraphics[width=1.0\linewidth]{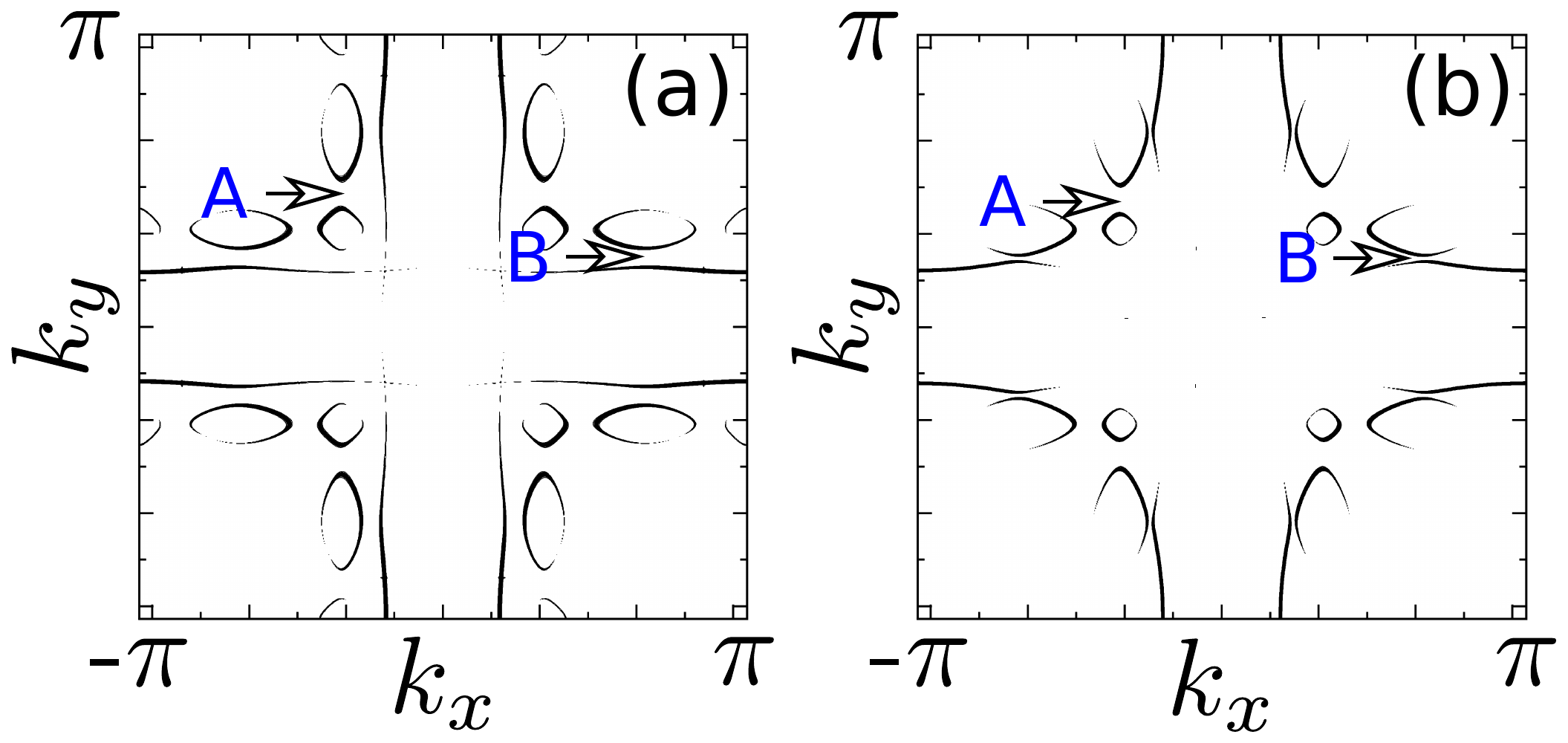} 
\caption{The spectral function in the biaxial CDW phase for the (a) Holstein and (b) B$_{1g}$ \textit{el-ph} coupling to the anti-bonding band at $10\%$ hole doping.} 
\label{fig:Fig_Holstein_biaxial}
\end{figure}

 As before, we start with the downfolded LDA dispersion from Ref.~\onlinecite{ANDERSEN19951573}. The energy spectrum and the corresponding eigenvectors are obtained by diagonalizing the Hamiltonian in Eq.~\ref{appeq.3}. The CDW band structure and the FS are shown in Fig.~\ref{fig:Fig3}a and Fig.~\ref{fig:Fig3}c, respectively, for $p=0.10$. Because of the commensurate period of three unit cells in each axial direction, the spectrum in Fig.~\ref{fig:Fig3}a contains nine bands in the RBZ. The five bands far away from the Fermi energy (illustrated as the solid lines) do not contribute to the FS reconstruction. The other four bands (enumerated as $\varepsilon_i, i=1,\ldots,4$) cross the Fermi energy and therefore determine the reconstructed FS. The lowest ($\varepsilon_1$) of these four bands yields one hole pocket around $(\pi/3,0)$, whereas the top-most band $\varepsilon_4$ leads to an electron pocket at the corner point $(\pi/3,\pi/3)$ of the RBZ. The remaining two intermediate bands $\varepsilon_{2,3}$ yield the open FS sheets shown in Fig.~\ref{fig:Fig3}c. The DOS $\mathcal{N}(\varepsilon)$ [Fig.~\ref{fig:Fig3}b] contains additional van Hove peaks and dips, in comparison to the uniaxial case. The origin of such peak and dip structure is again connected the reconstructed band structure at the relevant filling fractions~\cite{PhysRevB.97.125147}.
 
The spectral function at the Fermi energy is shown in Fig.~\ref{fig:Fig_Holstein_biaxial}.  As before, the influence of the matrix element $\mathsf{g}(\vq^\ast; \vk)$ is seen through a comparison of the B$_{1g}$ phonon to the Holstein model.  As in Sec.~\ref{sec:quality}, we adjust the strength of the Holstein coupling so that the band splittings at hotspot A are the same. Both Fermi surfaces have the same topology as in Fig.~\ref{fig:Fig3}, but two important distinctions are again clear:  the FS backfolding is far less apparent in the B$_{1g}$ phonon model than in the Holstein model, and the band splitting at hotspot B is much smaller, by a factor of 5, in the B$_{1g}$ phonon model. The distinction between the two models is again due to the strong momentum dependence of the B$_{1g}$ \textit{el-ph} matrix element.

\subsection{Quantitative estimates \label{sec:estimate_biaxial}}

We proceed as in Sec.~\ref{sec:estimate}. As for the uniaxial case, the FS reconstruction generated by the CDW is not readily apparent in the spectral function (Fig.~\ref{fig:Fig4}b), obtained from Eq.~\ref{appeq.3}, and  the band splittings at  hotspots A and B take the same values  as in the uniaxial case (21~meV and 3.6~meV, respectively). 

Whereas the uniaxial CDW generates a single hole pocket, the biaxial CDW generates a small diamond-shaped electron pocket, centered around the corners of the RBZ at $(\pm \pi/3, \pm \pi/3)$, and two hole pockets centered on the faces of the RBZ at $(\pm \pi/3, 0)$ and $(0, \pm \pi/3)$. The electron pocket in Fig.~\ref{fig:Fig4}a, has grown in size in comparison to Fig.~\ref{fig:Fig3}c because of the smaller hotspot gap at B, and has an area $A \sim 0.2/a_0^2$. This provides a frequency $F_{\text{elec}} = (\hbar c/2\pi e) A \sim 135$ T. In contrast, the hole pocket centered around $(\pi/3,0)$ has the larger area $A \sim 0.6/a_0^2$ which translates to the QO frequency $F_{\text{hole}} \sim 410$ T. Prominent quantum oscillations in YBa$_2$Cu$_3$O$_{6.5}$, in combination with the negative Hall coefficient, have been attributed to the electron pocket~\cite{Sebastian_2012,luoista}, but evidence for the hole pocket is scarce.  There is a single report of quantum oscillations with a frequency that is consistent with the predicted hole pocket\cite{Doiron-Leyraud2015}; however, the amplitude of these oscillations is very weak, roughly one twentieth of those for the electron pocket. There is no clear reason for this, although enhanced quasiparticle scattering on the hole pocket is frequently invoked as a plausible explanation.

\begin{figure}[t]
\centering
\includegraphics[width=1.0\linewidth]{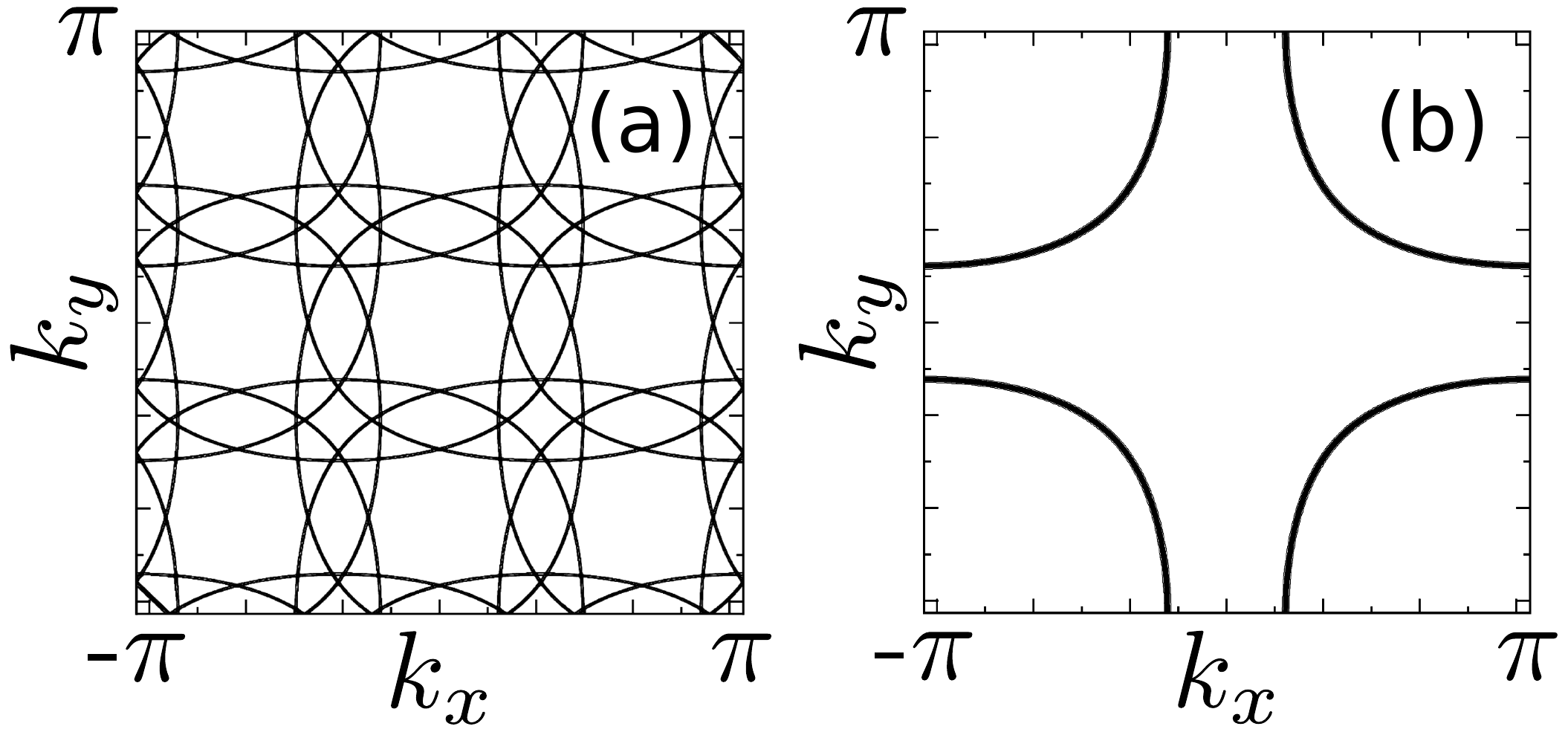} 
\caption{(a) The CDW Fermi surface for $10 \%$ hole doping obtained from the Hamiltonian in Eq.~\ref{appeq.3} with $\Delta_0 \sim 14$ meV~\cite{Forgan2015} and fitted dispersion $\varepsilon^{\text{R}}_\vk$~\cite{PhysRevB.75.184514}. (b) The spectral function in the CDW phase for the same parameters.} \label{fig:Fig4}
\end{figure}

Here, we point to an additional mechanism that may play a role.  Namely, quantum oscillations may be reduced because of enhanced magnetic breakdown for the hole pockets.  Figure~\ref{fig:Fig_Holstein_biaxial} shows that electrons orbiting the hole pocket under the influcence of the magnetic field pass by avoided Fermi surface crossings at hotspots of type A (separating electron and hole pockets) and type B (separating hole pockets and open Fermi surface sheets).  The term "magnetic breakdown" refers to the process by which a fraction $Q = e^{-B_0/B}$ of electrons jumps between Fermi surface branches as they move past each hotspot. In this expression, $B$ is the magnetic field and $B_0$ is the magnetic breakdown field,\cite{sebastian_quantum_2012,Shoenberg_1985}
\begin{equation}
B_0 \approx \frac{m^\ast}{e\hbar} \frac{\varepsilon_g^2}{\varepsilon_{\mathrm{F}}},
\end{equation}
where $\varepsilon_{\mathrm{F}}$ is the Fermi energy and $\varepsilon_g$ is the band splitting at the relevant hotspot.

Based on quantitative estimates, we suggested in Sec.~\ref{sec:estimate} that $\varepsilon_{g} = 21$~meV at hotspot A, and it is approximately 6 times smaller at hotspot B.  This distinction is tied to the strong momentum selectivity of the B$_{1g}$ phonon matrix element, and leads to a factor of 36 difference in the values of $B_0$ at the two hotspots.  Thus, it is possible that while $Q \ll 1$ for the electron pocket, such that there is very little reduction of the quantum oscillation amplitude, $Q$ can be of order 1 for electrons near hotspot B on the hole pocket.  This would lead to a strong suppression of the quantum oscillation signal for the hole pocket, proportional to $(1-Q)^2$ (as there are two hotspots of type B for each hole pocket). However, we caution that the equation for $Q$ predicts a magnetic field dependence for the oscillation amplitude that is not evident in the limited data that is currently available.

Finally, we comment on the symmetry of the charge modulation for biaxial order.  To a first approximation we expect the two components of the CDW, with wavevectors $\vq^*$ and $\overline{\vq}^*$, to be independent of each other.  This is because the Fermi surface hotspots corresponding to these components are far apart, and do not interfere with each other to first order in the CDW potential.  The symmetry components ($A_s$, $A_{s'}$, and $A_d$) are therefore expected to be the same for each of the ordering wavevectors as for the uniaxial case.  Higher-order terms, in which quasiparticles are scattered multiple times by the CDW, will lead to a mixing of the two CDW components and will determine whether or not biaxial order is energetically preferred to uniaxial order; however, pinning by disorder will also play an important role and will nucleate both CDW orientations.

\section{Conclusion \label{sec:conclu}}

In this paper, we analyzed the properties of the CDW phase in underdoped YBCO as if they emerge from a phonon-based mechanism~\cite{banerjee2019phononinduced} triggered by a unique cooperation of the B$_{1g}$ bond-buckling phonon and the quasiparticle renormalization due to strong electronic correlations. While the strength of the \textit{el-ph} coupling appears insufficient to generate long-range charge order, the experimental observations of medium-range charge correlations with incommensurate uniaxial ordering wave vectors near $0.3$~\cite{Ghiringhelli821,Chang2012,PhysRevLett.110.137004} along with considerable phonon softening~\cite{PhysRevLett.107.177004} support such a phonon based origin. We analyzed the reconstruction of the FS for both uni- and biaxial charge order. With only one free input  parameter \--- the static lattice distortion $\xi_\vq$, we contrasted the formation of a hole pocket in the uniaxial CDW FS to the formation of electron and hole pockets in the biaxial CDW FS. Specific for the \textit{el-ph} coupling of the B$_{1g}$ phonon is the result that the charge in the unit cell is redistributed with a predominant $s$-wave form factor.

Without aiming for quantitative accuracy, we nevertheless translate our results to various estimates. We have fixed $\xi_\vq$ in our ansatz to the experimental value~\cite{Forgan2015} for YBCO and adopted the phenomenological dispersion $\varepsilon^{\text{R}}_\vk$, to obtain the following quantum oscillation frequencies as 
\begin{equation}\label{eq.prediction}
F = \left\{
\begin{array}{@{}rl@{}}
& 1260 \; \text{T}, \quad \quad \quad \; \; \;  \text{uniaxial -- hole pocket}\\
& F_{\text{elec}} = 135 \; \text{T}, \quad   \text{biaxial -- electron pocket}\\
& F_{\text{hole}} = 410 \; \text{T}, \quad   \text{biaxial -- hole pocket}.
\end{array}
\right . 
\end{equation}
The origin of the observed central frequency $F_{\text{expt}} \sim 530$ T in underdoped YBCO~\cite{Doiron-Leyraud2007,PhysRevLett.104.086403,Riggs2011}, has been previously argued~\cite{Sebastian_2012} to be related to the diamond-shaped electron pockets (centered at the RBZ corner $(\pi/3,\pi/3)$ in Fig.~\ref{fig:Fig4}a). 

We note that the discrepancy between the predicted QO frequencies and experiments is significant. However, it does not represent a failure of the B$_{1g}$ phonon based mechanism.  Rather, the QO frequencies depend primarily on the band structure and the ordering wavevector $\vq^*$.  In Ref.~\onlinecite{Atkinson_2015}, it was shown that small changes in the band structure can easily change the QO frequency of the electron pocket by a factor of 3, which would bring our estimates within the experimental range.

Moreover, the area of the electron pocket decreases with increasing hole doping and hence the relevant frequency $F_{\text{elec}}$ should drop with hole doping. In contrast, $F_{\text{hole}}$ increases with hole doping and lies within the scale of the experimental data~\cite{PhysRevB.59.14618}. Our results have also led to a magnetic breakdown scenario for the hole-pocket orbits in the biaxial CDW state. This breakdown offers an explanation to why the corresponding QO frequency is barely observed.

With a predominant $s$-wave character~\cite{mcmahon2019orbital} of the charge modulation, we obtain an amplitude of approximately $0.003e$ for the variation of the charge on the copper atom, which is ten times smaller than the experimental estimates from $^{63}$Cu NMR lines~\cite{Wu2011}. Our smaller charge modulation amplitude, however, comes closer to the result of the NQR linewidth analysis by Kharkov and Sushkov~\cite{Kharkov2016}.

The smallness of the static B$_{1g}$ lattice distortion $\xi_\vq$ and the associated CDW energy scale $\Delta_0 = 14$ meV have two implications. First, the FS reconstruction due to uni- or bi-axial charge order is rendered barely observable for ARPES experiments. Second, the CDW energy scale is smaller but comparable to the superconducting $d$-wave gap parameter $\Delta_{\mathrm{sc}} \sim 50$ meV in underdoped cuprates. Charge order and superconductivity compete with respect to their individual free-energy gain controlled by either $\Delta_0$ or $\Delta_{\mathrm{sc}}$. In the doping range where charge order exists in the phase diagram of YBCO, the superconducting $T_{\mathrm{c}}$ is little suppressed~\cite{Comin}. And upon entering the superconducting state via cooling through $T_{\mathrm{c}}$, CDW x-ray intensities~\cite{Ghiringhelli821,Chang2012,PhysRevLett.110.137004,PhysRevLett.110.187001,PhysRevLett.109.167001} and CDW correlation length shrink significantly~\cite{Comin}. Both observations are expected and compatible with our finding $\Delta_0 = 14$ meV $< \Delta_{\mathrm{sc}} = 50$ meV.

Throughout this paper, we have primarily focused on YBCO; however, we believe that a similar mechanism may apply to Bi-based cuprates, for example Bi$_2$Sr$_2$CaCu$_2$O$_{8+x}$.  There, the analysis is complicated by the presence of a lattice supermodulation with a period of approximately 5 unit cells~\cite{PhysRevB.77.220507} that may interfere with the CDW. A detailed analysis of such effects is beyond the scope of this work.

 Our results support the notion that charge order in underdoped cuprates is not primarily caused by a purely electronic mechanism. Rather, pursuing the assumption of a B$_{1g}$ type lattice distortion leads to a CDW characteristics compatible with experimental observations.

\begin{acknowledgments}
WAA acknowledges support by the Natural Sciences and Engineering Research Council (NSERC) of Canada.  This work is supported by the Deutsche Forschungsgemeinschaft (DFG, German Research Foundation)- project-ID-107745057-TRR 80.
\end{acknowledgments}

\appendix

\section{Anti-bonding band: dispersions and eigenfunctions \label{sec:downfolding}}

We start from the downfolded three-band Hamiltonian $\text{H}_{\text{kin}}$ as
\begin{equation}\label{down}
\scalebox{0.97}[1]{$\text{H}_{\text{kin}} = \sum_{\vk}\bm{\Psi^{\dagger}_{\vk}}
\begin{pmatrix}
\varepsilon_d  & 2t_{pd}s_x    & -2t_{pd}s_y\\
    2t_{pd}s_x     & \tilde{\varepsilon}_x(\vk) & 4\tilde{t}_{pp}s_x s_y \\
    -2t_{pd}s_y    & 4\tilde{t}_{pp}s_xs_y & \tilde{\varepsilon}_y(\vk)
\end{pmatrix} \bm{\Psi}_{\vk},$}
\end{equation}
with the three-spinor $\bm{\Psi^{\dagger}_{\vk}} = \left( d^{\dagger}_{\vk},\; p^{\dagger}_{x\vk},\; p^{\dagger}_{y \vk}\right)$ and $s_{x,y} = \sin(k_{x,y}/2)$. $\varepsilon_d$ denotes the onsite energy of the $d$ orbital, $t_{pd}$ is the hopping amplitude between $p$- and $d$-orbitals. In the downfolding procedure, the hopping processes via the copper $4s$ orbital renormalize the oxygen energies $\varepsilon_p$ and generate indirect hopping $4t^i_{pp}$ between oxygen orbitals:
\begin{equation}\label{Onsite}
\tilde{\varepsilon}_{x,y} = \varepsilon_p + 4t^i_{pp} s^2_{x,y};\; \tilde{t}_{pp} = t^i_{pp} + t^d_{pp};\;  t^i_{pp} = \frac{t^2_{ps}}{\varepsilon_{\mathrm{F}}-\varepsilon_s},
\end{equation}
where $\varepsilon_{\mathrm{F}}$ is the Fermi energy, $t^d_{pp}$ a small direct hopping amplitude, and $t_{ps}$ denotes the hopping amplitude between $d$- and $s$-orbitals. We adopt all the parameters entering Eqs.~\ref{down} and \ref{Onsite} from Ref.~\onlinecite{ANDERSEN19951573}, specifically $t_{pd} = 1.6$ eV, $\varepsilon_d-\varepsilon_p = 0.9$ eV, $t^d_{pp} = 0$ and $t^i_{pp} = -1.0$ eV. We diagonalize  $\text{H}_{\text{kin}}$ and focus on the only partially filled anti-bonding band; the irrelevant spin index is suppressed. We obtain the dispersion for the anti-bonding band as
\begin{align}\label{anti-bonding}
\varepsilon_\vk  = & \frac{\varepsilon_d+\tilde{\varepsilon}_{x}+\tilde{\varepsilon}_{y}}{3} + 2 \mathrm{Re} \left(A_{\vk} + \sqrt{A_{\vk}^2+B_{\vk}^3} \right)^{\frac{1}{3}},
\end{align}
where the parameters $A_{\vk}$ \& $B_{\vk}$ are defined as
\begin{widetext}
\begin{align}\label{parameter}
A_{\vk} & = \frac{\varepsilon_d}{6}\left( t_x^2 +t_y^2 -2t'^2\right) +\frac{\tilde{\varepsilon}_x}{6}\left( t'^2 +t_x^2 -2t_y^2\right) +\frac{\tilde{\varepsilon}_y}{6}\left( t'^2 +t_y^2 -2t_x^2\right) - t't_xt_y\\ \nonumber
&  + \frac{\varepsilon_d^3+\tilde{\varepsilon}_x^3+\tilde{\varepsilon}_y^3}{27}-\frac{\varepsilon_d^2\tilde{\varepsilon}_x+\tilde{\varepsilon}_x^2\varepsilon_d+\tilde{\varepsilon}_x^2\tilde{\varepsilon}_y+\tilde{\varepsilon}_y^2\tilde{\varepsilon}_x+\varepsilon_d^2\tilde{\varepsilon}_y+\tilde{\varepsilon}_y^2\varepsilon_d}{18} + \frac{2\varepsilon_d\tilde{\varepsilon}_x\tilde{\varepsilon}_y}{9} \\ \nonumber
B_{\vk} &= -\frac{\varepsilon_d^2+\tilde{\varepsilon}_x^2+\tilde{\varepsilon}_y^2-\varepsilon_d\tilde{\varepsilon}_x-\tilde{\varepsilon}_x\tilde{\varepsilon}_y-\tilde{\varepsilon}_y\varepsilon_d}{9}-\frac{1}{3} \left( t_x^2 +t_y^2 + t'^2 \right),
\end{align}
where $t_x=2t_{pd}s_x$, $t_y = 2t_{pd}s_y$, and $t'=4\tilde{t}_{pp}s_xs_y$. The projections of $d$, $p_x$ and $p_y$ orbitals onto the anti-bonding band follow from the eigenfunctions for the anti-bonding band as 
\begin{align}\label{eigenfunction}
\phi_d(\vk)  & = \frac{1}{N_\vk} \Big[ (\varepsilon_\vk-\tilde{\varepsilon}_x)(\varepsilon_\vk-\tilde{\varepsilon}_y) - t'^2 \Big], \; \phi_x(\vk) = \frac{1}{N_\vk} \Big[ (\varepsilon_\vk-\tilde{\varepsilon}_y) t_x - t't_y \Big], \; \phi_y(\vk)   = -\frac{1}{N_\vk} \Big[ (\varepsilon_\vk-\tilde{\varepsilon}_x) t_y - t't_x \Big],\\ \nonumber
& \qquad N_\vk  = \sqrt{\Big[ (\varepsilon_\vk-\tilde{\varepsilon}_x)(\varepsilon_\vk-\tilde{\varepsilon}_y) - t'^2 \Big]^2 + \Big[ (\varepsilon_\vk-\tilde{\varepsilon}_y) t_x - t't_y \Big]^2 + \Big[ (\varepsilon_\vk-\tilde{\varepsilon}_x) t_y - t't_x \Big]^2}.
\end{align}
\end{widetext}
We notice that $(A_{\vk}^2+B_{\vk}^3)$ is negative for all $\vk$ in the Brillouin zone. Hence, we can write Eq.~\ref{anti-bonding} as
\begin{equation}\label{anti-bonding_Extra}
\varepsilon_\vk  = \frac{\varepsilon_d+\tilde{\varepsilon}_{x}+\tilde{\varepsilon}_{y}}{3} + 2\text{Re} \Big[\left(A_{\vk} + \sqrt{A_{\vk}^2+B_{\vk}^3} \right)^{\frac{1}{3}} \Big].
\end{equation}
A simplified version of Eq.~(\ref{eigenfunction}) and Eq.~(\ref{anti-bonding_Extra}) was obtained in Refs.~\onlinecite{PhysRevB.59.14618,PhysRevLett.93.117004}. 

\section{Biaxial CDW: Reconstruction \label{sec:bidirection}}

In this Appendix, we provide the details of the mean-field Hamiltonian described in Sec.~\ref{sec:bidir}. The quasiparticles in the biaxial CDW state are described by a 9-component spinor as
\begin{equation}\label{appeq.1}
\mathcal{H}_{\text{MF}} = \sum_{\vk \in \text{RBZ}} \Psi^{\dagger}_\vk \underset{\sim}{\text{H}}(\vk) \Psi_\vk,
\end{equation}
where the wave vector $\vk$ belongs to the RBZ determined by by the restriction $|k_x|,|k_y| \le \pi/3$. The nine-component spinor $\Psi_\vk$ is defined as
\begin{widetext}
\begin{align}\label{appeq.2}
\Psi_\vk  = & ( c_\vk, c_{\vk+\vq^*}, c_{\vk-\vq^*}, c_{\vk+\overline{\vq}^*}, c_{\vk+\vq^* +\overline{\vq}^*}, c_{\vk-\vq^* +\overline{\vq}^*}, c_{\vk-\overline{\vq}^*}, c_{\vk+\vq^* -\overline{\vq}^*}, c_{\vk-\vq^* -\overline{\vq}^*} )^{\text{T}}.
\end{align}
The Hamiltonian matrix $\underset{\sim}{\text{H}}(\vk)$ is obtained (with the order parameter $\Delta(\vq^*;\vk)$ defined in Sec.~\ref{sec:unidir}) as 
\begin{equation}\label{appeq.3}
\scalebox{0.578}[1]{$\underset{\sim}{\text{H}}(\vk) = \begin{pmatrix}
\varepsilon_\vk & \Delta(\vq^*;\vk) & \Delta(\vq^*;\vk- \vq^*) & \Delta(\overline{\vq}^*;\vk) & 0 & 0 & \Delta(\overline{\vq}^*;\vk-\overline{\vq}^*) & 0 & 0 \\  
\Delta(\vq^*;\vk) & \varepsilon_{\vk+\vq^*} & \Delta(\vq^*;\vk+ \vq^*) & 0 & \Delta(\overline{\vq}^*;\vk+\vq^*) & 0 & 0 & \Delta(\overline{\vq}^*;\vk+\vq^*-\overline{\vq}^*) & 0 \\  
\Delta(\vq^*;\vk- \vq^*) & \Delta(\vq^*;\vk+ \vq^*) & \varepsilon_{\vk-\vq^*} & 0 & 0 & \Delta(\overline{\vq}^*;\vk-\vq^*) & 0 & 0 & \Delta(\overline{\vq}^*;\vk-\vq^*-\overline{\vq}^*) \\  
\Delta(\overline{\vq}^*;\vk) & 0 & 0 & \varepsilon_{\vk+\overline{\vq}^*} & \Delta(\vq^*;\vk+ \overline{\vq}^*) & \Delta(\vq^*;\vk - \vq^* +\overline{\vq}^*) & \Delta(\overline{\vq}^*;\vk+ \overline{\vq}^*) & 0 & 0 \\  
0 & \Delta(\overline{\vq}^*;\vk+\vq^*) & 0 &  \Delta(\vq^*;\vk+ \overline{\vq}^*) & \varepsilon_{\vk+\vq^*+\overline{\vq}^*} & \Delta(\vq^*;\vk+ \vq^* +\overline{\vq}^*) & 0 & \Delta(\overline{\vq}^*;\vk+\vq^*+\overline{\vq}^*) & 0 \\  
0 & 0 & \Delta(\overline{\vq}^*;\vk-\vq^*) & \Delta(\vq^*;\vk - \vq^* +\overline{\vq}^*) & \Delta(\vq^*;\vk+ \vq^* +\overline{\vq}^*) & \varepsilon_{\vk-\vq^* +\overline{\vq}^*} & 0 & 0 & \Delta(\overline{\vq}^*;\vk-\vq^*+\overline{\vq}^*) \\  
\Delta(\overline{\vq}^*;\vk-\overline{\vq}^*) & 0 & 0 & \Delta(\overline{\vq}^*;\vk+ \overline{\vq}^*) & 0 & 0 & \varepsilon_{\vk-\overline{\vq}^*} & \Delta(\vq^*;\vk- \overline{\vq}^*) & \Delta(\vq^*;\vk- \vq^* -\overline{\vq}^*) \\  
0 & \Delta(\overline{\vq}^*;\vk+\vq^*-\overline{\vq}^*) & 0 & 0 & \Delta(\overline{\vq}^*;\vk+\vq^*+\overline{\vq}^*) & 0 & \Delta(\vq^*;\vk- \overline{\vq}^*) & \varepsilon_{\vk+\vq^* -\overline{\vq}^*} & \Delta(\vq^*;\vk+ \vq^* -\overline{\vq}^*) \\  
0 & 0 & \Delta(\overline{\vq}^*;\vk-\vq^*-\overline{\vq}^*) & 0 & 0 &  \Delta(\overline{\vq}^*;\vk-\vq^*+\overline{\vq}^*) & \Delta(\vq^*;\vk- \vq^* -\overline{\vq}^*) & \Delta(\vq^*;\vk+ \vq^* -\overline{\vq}^*) & \varepsilon_{\vk - \vq^* -\overline{\vq}^*}  
\end{pmatrix}$}.
\end{equation}
\end{widetext}

\bibliographystyle{apsrev4-1}
\bibliography{Charge_cuprates}

\end{document}